\documentclass[12pt, draftclsnofoot, onecolumn]{IEEEtran} 
 \usepackage{amsmath,amssymb}
 \usepackage{subfigure}
 \usepackage{graphicx,graphics,psfrag}
 \usepackage{cite,balance}
 \usepackage{algorithm}
 \usepackage{accents}
 \usepackage{amsthm}
 \usepackage{bm}
 \usepackage{url}
 \usepackage{algorithmic}
 \usepackage[english]{babel}
 \usepackage{multirow}
 \usepackage{enumerate}
 \usepackage{cases}
 \usepackage{stfloats}
 \usepackage{dsfont}
 \usepackage{soul}
 \usepackage{amsfonts}
\usepackage[utf8]{inputenc}
 \usepackage{cite,graphicx,amsmath,amssymb}
 \usepackage{subfigure}
 \usepackage{fancyhdr}
 \usepackage{hhline}
 \usepackage{graphicx,graphics}
 \usepackage[dvipsnames]{xcolor}

\IEEEoverridecommandlockouts

\begin{document}

\title{Task-Oriented Integrated Sensing, Computation and Communication for Wireless Edge AI}
\author{Hong Xing, Guangxu Zhu, Dongzhu Liu, Haifeng Wen, Kaibin Huang, and Kaishun Wu
\thanks{H. Xing and H. Wen are with the Internet of Things (IoT) Thrust, The Hong Kong University of Science and Technology (Guangzhou), Guangzhou 511453, China. H. Xing is also affiliated with the Department of Electronic and Computer Engineering, The Hong Kong University of Science and Technology, Hong Kong (e-mails: hongxing@ust.hk, hwen904@connect.hkust-gz.edu.cn).}
\thanks{G. Zhu is with Shenzhen Research Institute of Big Data, Shenzhen 518172, China  (e-mail: gxzhu@sribd.cn).}
\thanks{D. Liu is with the School of Computing Science, University of Glasgow, Glasgow G12 8RZ, United Kingdom (e-mail: dongzhu.liu@glasgow.ac.uk).}
\thanks{K. Huang is with the Department of Electrical and Electronic Engineering (EEE), The University of Hong Kong, Hong Kong (e-mail: huangkb@eee.hku.hk).}
\thanks{K. Wu is with the Data Science and Analytics (DSA) Thrust and the Internet of Things (IoT) Thrust, The Hong Kong University of Science and Technology (Guangzhou), Guangzhou 511453, China, and is also with the College of Computer Science and Software Engineering, Shenzhen University, Shenzhen 518066, China (e-mail: wuks@ust.hk).}
}

\markboth{}{}
\maketitle

\setlength\abovedisplayskip{2pt}
\setlength\belowdisplayskip{2pt}

\vspace{-1.5cm}

\begin{abstract}
With the advent of emerging IoT applications such as autonomous driving, digital-twin and metaverse etc. featuring massive data sensing, analyzing and inference as well critical latency in \emph{beyond 5G (B5G)} networks, \emph{edge artificial intelligence (AI)} has been proposed to provide high-performance computation of a conventional cloud down to the network edge. Recently, convergence of wireless \emph{sensing, computation and communication (SC${}^2$)} for specific edge AI tasks, has aroused paradigm shift by enabling (partial) sharing of the radio-frequency (RF) transceivers and information processing pipelines among these three fundamental functionalities of IoT. However, most existing design frameworks separate these designs incurring unnecessary signaling overhead and waste of energy, and it is therefore of paramount importance to advance fully \emph{integrated sensing, computation and communication (ISCC)} to achieve ultra-reliable and low-latency edge intelligence acquisition. In this article, we provide an overview of principles of enabling ISCC technologies followed by two concrete use cases of edge AI tasks demonstrating the advantage of task-oriented ISCC, and pointed out some practical challenges in edge AI design with advanced ISCC solutions. 
\end{abstract}

%

\section{Introduction}\label{sec:Introduction}
With the proliferation of massive data generated by IoT devices and their applications such as smart transport, smart city and digital twin etc., there is unprecedented demand for the network edge to be capable of contextual awareness and extensive computation, thus achieving ultra reliable and low-latency intelligence acquisition. To satisfy such requirements, wireless edge \emph{artificial intelligence (AI)} has been proposed as a promising solution to pull cloud functionalities including high-performance computation, communications and control down to the proximity of the IoT front end \cite{MChenJSAC_ML}. By generating and consuming data locally, and based upon it performing task-oriented training and inference all within the network edge, edge AI solution significantly reduces energy and end-to-end latency for network services, thus being one of the key enablers for beyond 5G (B5G)  IoT with ubiquitous perception and endogenous intelligence.

Among a variety of design components, (wireless) sensing, computation and communication are three building blocks that are tightly coupled each other for effective edge AI. Wireless sensing enabled by contactless sensors features non-intrusiveness, reliability, scalability and resilience, and is thus widely adopted to support various ambient intelligence applications for, e.g., health monitoring and traffic surveillance \cite{wu2017WiFall}. To exploit the wealth of sensing data to provide services like target detection, positioning and motion recognition, it is imperative to first analyze these raw signals by proper filtering and time-frequency analysis, and then feed the extracted features into the input of the AI model to collectively train a shared AI model or make device-edge co-inference, meanwhile providing reliable and low-latency communications between the edge server and edge devices that may involve model/gradient parameters in million-to-billion order of magnitudes. For example, in vehicular-to-everything (V2X) networks, a unified waveform bearing both vehicle-to-road side unit (RSU) information and sensing signal is expected to attain simultaneous high data-rate communication and broad field of view beyond the vehicle's line-of-sight. Meanwhile, real-time processing of massive sensed data for collaborative AI model inference among RSUs is also favored to assist with safe driving. However, on one hand, it may be infeasible to deploy such complicated DNN on edge devices for on-device inference using limited on board computation and storage resources. On the other hand, offloading the data to the cloud for remote inference may violate the latency requirement and incur terrible accidents.
These dilemmas, needs and opportunities facilitate the development of technologies such as wireless edge learning/inference, and furthermore, motivate joint \emph{sensing, computation, and communication (SC${}^2$)} designs.

Most of existing SC${}^2$ solutions focus on joint design of any two factors out of the three. Recently, the wireless sensing and communication layer tend to converge into a novel \emph{signaling layer} in the state-of-the-art (SOTA) IoT architecture \cite{cui2021ISAC}. In most prior arts, radio-frequency (RF) sensing and/or communication performed at the signaling layer and AI model training/inference employed at the upper application layer are often two processes that are separately designed, which hardly achieves uniformly satisfactory performance across different edge AI tasks due to, e.g., mismatch of resources. For instance, if low transmit power is tuned by battery-limited wireless sensors to generate some low-resolution data that are just sufficient for anomaly detection, but later these data samples are used for high-complex model training, then a waste of computation energy is inevitable, let alone possibly poor generality performance due to over-fitting. To streamline wireless edge AI deployment with significantly improved energy efficiency and reduced hardware as well as signaling cost, it is critical to breaking boundaries among SC${}^2$ in a conventional design framework that separates them as three relatively independent modules --- allowing sensed information to enhance the reliability of communication and computation, and computation performance to guide through sensing and computation in turn. 

While \cite{Li2023ISCCair} proposed a shared framework of \emph{integrated sensing, computation and communication (ISCC)} over the air, in which multiple multi-antenna IoT sensors transmit signals for simultaneous target detection and over-the-air data fusion, and jointly optimized SC${}^2$ beamformers to improve trade-offs between their mean squared errors (MSE). However, there is still a lack of systematic understanding of ISCC in literature, in particular, implementations that are capable of providing closed-loop data and control flow across the SC${}^2$ processes with feedback and adaption to tasks except the first few attempts aiming for harnessing their mutual benefits in edge AI tasks \cite{liu22ISCC,Wen-TWC2022}. As a result, this paper advocates a design framework of fully ISCC to unleash the full potential of edge AI for ultra-reliable and low-latency intelligence acquisition.

The structure of the article is organized as follows. Section~\ref{sec: Enabling Technologies} presents fundamental principles and advantages of three enabling technologies for ISCC. The commonly adopted performance metrics for edge AI tasks along with concrete case studies of human motion recognition factoring ISCC are elaborated in Section~\ref{sec: Task-Oriented ISCC for Wireless FL} and Section~\ref{sec: Task-Oriented ISCC for Wireless EI}, respectively. In Section~\ref{sec: Advanced Task-Oriented ISCC Techniques}, we identify some challenges in implementations of ISCC-based edge AI tasks with corresponding SOTA solutions. Finally, Section~\ref{sec: Conclusions} concludes the article with discussions on some open challenges in implementations of ISCC techniques.


\section{Enabling Technologies for Task-Oriented ISCC}\label{sec: Enabling Technologies}

\begin{figure}[t]
\centering
\setlength{\abovecaptionskip}{-1mm}
\setlength{\belowcaptionskip}{-1mm}
   \includegraphics[width=4.0in]{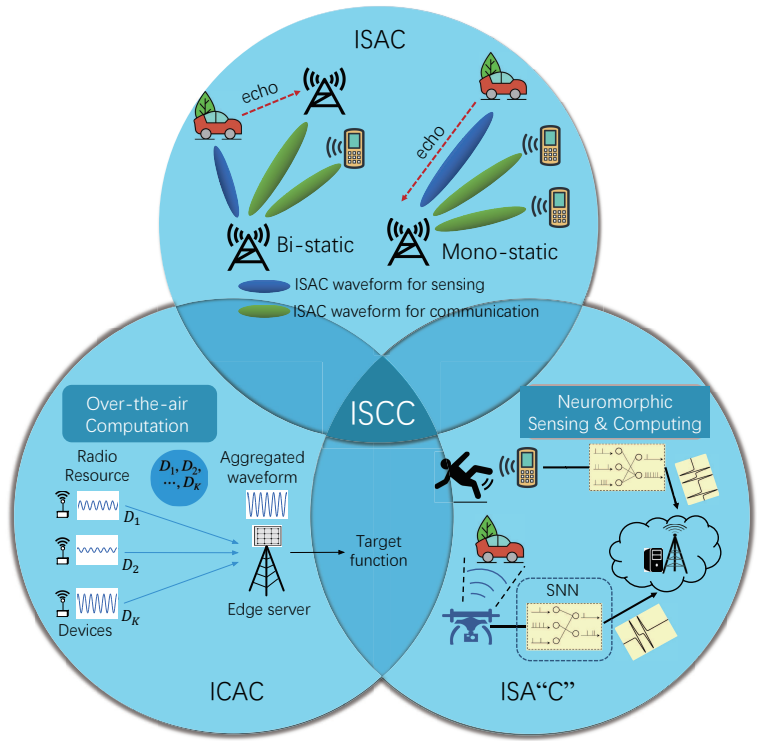}
\caption{The relationship diagram between ISCC and enabling technologies.} \label{fig:ISCC_relations}
\end{figure}


\subsection{Integrated Sensing and Communication (ISAC)}
With the growing demand for ubiquitous perception inherent to IoT services such as extended reality (XR), autonomous driving, and metaverse, etc., two fundamental functionalities of any radio-frequency (RF) systems, (wireless) sensing and communication, is experiencing paradigm shift from separate modules with limited hardware and resources sharing to an integrated architecture, namely, \emph{integrated sensing and communication (ISAC)}, with these two factors being jointly designed in terms of both hardware platform and information processing pipeline \cite{liu2020JCC}. 
The principle behind ISAC is that an RF waveform can serve both the purpose of conveying information from the transmitter (Tx) to the receiver (Rx), and of retrieving Doppler shift, angles of departure/arrival and range, etc. from its echoes. The basic ISAC-enabled system can be classified into mono-static and bi-static setup (c.f.~Fig. \ref{fig:ISCC_relations}), depending on whether the sensing information is estimated on a co-located or separate Rx. In spite of the ultimate goal of fully unified waveform for sensing and communications with minimal modification to existing communications infrastructure, a practical ISAC system may also partially implement the concept of ``ISAC'' by coordinating these two functionalities over some non-overlapping network resource, achieving partial \emph{integration gain (IG)}. IG refers to various advantages gained by sensing and communication being integrated to different extents, including spectral and energy efficiency, hardware cost and signaling latency reduction. For instance, if the knowledge of the surrounding radio environment inferred by the sensing Rx such as channel state information (CSI) can be accessed in a cross-function manner, the Tx can exploit such information to facilitate communication design, further achieving \emph{coordination gain (CG)} between sensing and communication in addition to IG.

\subsection{Integrated Computation and Communication (ICAC)}
Recent years have witnessed a rapid development of AirComp as a promising technology for low-latency data aggregation in IoT networks \cite{Zhu-WCM21}. The magic behind the technique is to exploit the superposition property of wireless channels for automatic data aggregation over the air, thus vividly turning the air into a computer. As opposed to the classic strategy of “communication before computing”, AirComp essentially integrates computing into communication, resulting in a new paradigm of \emph{integrated computation and communication (ICAC)}. As a result, unlike traditional wireless communication over a multi-access channel (MAC), which requires orthogonal transmission for successful decoding of individual messages from different devices, AirComp allows edge devices to simultaneously transmit their respective signals on the same frequency band with proper processing, such that the functional computation of the distributed data is accomplished directly over the air, as shown in Fig. \ref{fig:ISCC_relations}. This thus significantly improves the communication and computing efficiency with theoretically striking performance gain in terms of multi-access latency up to the network population, e.g., $100$ times faster for a network consisting of $100$ participating devices.

\subsection{Integrated Sensing and Computation (ISA``C'')}
As next-generation wireless networks evolve, many emerging applications require intelligent edge devices to sense their surroundings to obtain cognitive information to perform certain computing tasks efficiently, e.g., fall detection, health monitoring, etc. Hence, the demands for \emph{integrated sensing and computation (ISA``C'')} and high energy efficiency are growing. In this context, inspired by the brain, representative technologies, neuromorphic sensing and computing, have emerged that enable deep integration of IoT sensors with back-end computing architectures (neuromorphic processors), where sensing information will directly serve computation tasks compared to traditional digital communications (c.f.~Fig. \ref{fig:ISCC_relations}). State-of-the-art neuromorphic processors are typical of spiking neural networks (SNNs) \cite{roy2019towards}. SNNs process information encoded in the timing of spikes, as opposed to the conventional deep neural networks (DNNs) that process continuous real numbers. A spike characterizes a binary event, either $0$ or $1$.
A structure of SNN has an arbitrary topology that is formed by the interconnection of numerous neurons, where each neuron is active only when transmitting or receiving spikes and idle otherwise. 
Neurons receive timings of spikes as inputs and form a membrane potential by filtering and weighted-sum operations to compare with a threshold, thus deciding whether it is active or not, while neurons in DNNs are always active. Facilitated by emerging large-scale neuromorphic chips, SNN can be implemented on-chip to achieve extremely low energy consumption and fast computation. Specifically, each spike only costs a picojoule of energy, leading to energy-efficient edge AI. 

\textcolor{black}{To sum up, when pushing towards boundaries of any combined two factors in Fig. \ref{fig:ISCC_relations}, i.e., ISAC, ICAC, and ISA``C'', we highlight the ISCC synergy from the following aspects. First, sensing can exploit its knowledge of surroundings to facilitate ICAC through more accurate channel estimation; and meanwhile capture sporadic changes in the perceived environment that is otherwise idle, thus saving a significant amount of energy for IoT devices, exemplified by neuromorphic edge AI architecture. Secondly, computation endows ISAC with distributed edge intelligence by directly providing highly efficient task-oriented solutions like target detection and motion recognition, or by guiding through channel estimation and beam tracking etc. using AI-assisted design. Lastly, communication can transform shared wireless medium into an inherent computing engine or replace traditionally module-based transmission chain with an autoencoder separated by wireless channels, enabling AirComp-based data aggregation or ISA``C'' based end-to-end training, respectively, both making ultra low-latency connected intelligence possible.
}

\section{Task-Oriented ISCC for Wireless Federated Learning}\label{sec: Task-Oriented ISCC for Wireless FL}
This section introduces the commonly adopted performance metrics in ISCC-based wireless FL followed by a concrete use case of human motion recognition, illustrating how joint SC${}^2$ resource allocation is implemented in the design flow ISCC to enhance the learning performance.

\subsection{Performance Metrics}
The commonly used performance metrics in wireless FL include convergence rate, learning latency, and energy consumption. When enabled by ISCC design, they are modified as follows.

\begin{itemize}
\item {} {\bf Convergence rate}: Convergence rate characterizes how the learning error, exemplified by the optimality gap of the training loss and expected average norm square of the gradient, varies over global communication rounds. It is nevertheless intractable to find the exact expression of learning error for general AI training models, and therefore upper bounds on the optimality gap of the training loss or the expected average norm square of the gradient, depending on assumptions on smoothness and/or convexity of the local loss function, are widely adopted to facilitate quantifying the convergence rate in the literature of wireless FL \cite{MChenJSAC_ML}.

\item {} {\bf Learning latency}: Learning latency is defined as the wall clock time for the learning process to converge within a given accuracy. The learning latency generally refers to the number of global communication rounds times per-round latency \cite{Cao2021-FedAvg}. In the settings of ISCC-based wireless FL, the per-round latency consists of the sensing time (including pre-processing of raw sensing signals), computation latency (the number of local update iterations times per-iteration computation time) for local gradient/model updating and the communication time for local gradient/model uploading (uplink) and global model broadcasting (downlink).

\item {} {\bf Energy consumption}: Energy consumption corresponds to the sensing, computation, and communication energy required for the overall learning process. The energy for sensing is solely consumed by edge devices mainly for sensing signal transmission. The energy for local computation is also consumed by edge devices for local gradient/model updating. The energy for communication is generally consumed by both edge devices (for local gradient/model uploading) and the edge server (for global model downloading).

\end{itemize} 
\begin{figure}[t]
\centering
 \setlength{\abovecaptionskip}{0in}
\setlength{\belowcaptionskip}{0in}
    \includegraphics[width=5.2in]{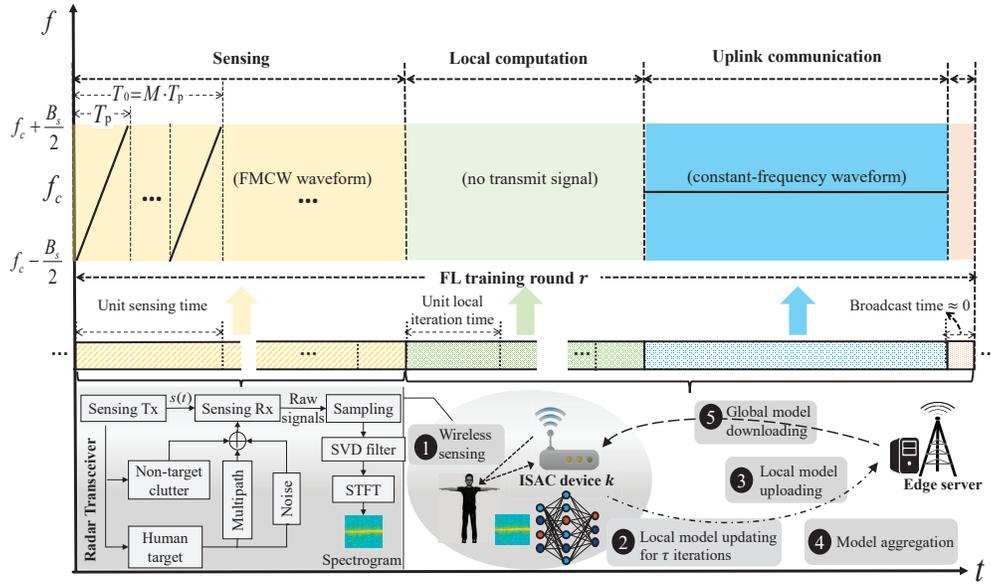}
  \caption{The diagram of complete pipelines of ISCC includes: 1) wireless sensing, 2) local model updating, 3) local model uploading, 4) global model aggregation, and 5) global model downloading, along with the time-frequency allocation within one communication round of FL training for human motion recognition.}
  \label{fig:pipeline of ISCC based FL}
\vspace{-0.1in}
\end{figure}

\subsection{\emph{Use Case: Human Motion Recognition}}
In this subsection, we provide a concrete use case of human motion recognition based on edge AI-enabled wireless sensing \cite{liu22ISCC}, demonstrating the advantage of ISCC-based resource allocation in terms of the trade-offs among the above-mentioned performance metrics. 

This use case considered ISCC-based resource allocation for a wireless FL system in which multiple ISAC devices obtain their respective datasets for human motion recognition by wireless sensing, and then exchange only model updates with the edge server. The time-frequency allocation of each ISAC device within one particular round of FL training is shown in Fig. \ref{fig:pipeline of ISCC based FL}. \textcolor{black}{The design flow primarily addressed two challenges in this setting: 1) how to establish between sensing and learning a closed-loop control flow, such that the performance of FL training can be analyzed offline and then guides through sensing process on-the-fly; 2) how to jointly optimize SC${}^2$ resource allocation, avoiding situations when mismatch between sensing quality and uplink transmission condition causes inefficient model aggregation at the edge server.} 

In this use case, sensing quality control can affect convergence rate through two factors, the $k$th device's sensing transmit power \(p_{s,k}\) and the number $b^{(r)}$ of samples generated at round $r$. As shown in Fig. \ref{fig:pipeline of ISCC based FL}, the input of the training model is the pre-processed spectrogram, and it was verified by experiments that the quality of a spectrogram will no longer improve as long as the sensing transmit power is large enough to combat the effects of the ground clutter and receiver noise. This finding suggests that maintaining sensing transmit power above a threshold value is sufficient to generate data samples of approximately the same satisfactory quality over the whole training process. Previous theoretical results showed that the convergence rate increases with the \emph{constant} per-round batch size under some other assumptions on hyperparameters. However, considering the task of human motion recognition, larger batch size of data samples indicates longer sensing time $T_0b^{(r)}$ and higher energy consumption $p_{s,k}T_0b^{(r)}$ in sensing at each round, which easily violates the total latency and energy constraints. This motivates \emph{adaptive} batch size design, i.e., proper choice of $b^{(r)}$ in each round to achieve optimal trade-off between the convergence rate and the task's latency and energy requirement. This bridges the gap between the ultimate learning performance and sensed data quality, and thus addresses Challenge 1).

As shown in Fig. \ref{fig:pipeline of ISCC based FL}, since the increase in the number of sensed data samples causes longer time as well as larger energy consumption in local computation and correspondingly quite limited model uploading time, thus imposing a higher requirement on the uplink transmission rate, benefits of accelerated convergence may be compromised by intolerably higher energy consumption or infeasible uplink transmit power. It is therefore imperative to address Challenge 2). In this use case, sensing and communication transmit power at each ISAC device, time for sensing and communication, and adaptive batch size at each round were jointly optimized to maximize the converge rate under practical transmit power, total latency and energy constraints. The formulated non-convex problem was solved by being decoupled into two subproblems. The first subproblem aims for maximizing the total number $\sum_rb^{(r)}$ of data samples generated in the training process by optimal joint SC${}^2$ resource allocation. Specifically, this subproblem problem was equivalently transformed into another problem with relevance to only the communication transmit power of each ISAC device, which can be further decoupled into the same number of small-scale problems as the number of ISAC devices, with each being single-variable and then solved at each device in parallel. The second subproblem aims for maximizing the convergence rate by partitioning the optimized total number of data samples in the first subproblem into $b^{(r)}$ at each round. To solve this problem with intractable objective function due to adaptive batch size, the authors in \cite{liu22ISCC} built upon the previous theoretical results, which suggests that the optimal batch size is in approximately linear proportion to the decrease in the value of loss function, a suboptimal closed-form solution for the batch size $b^{(r)}$. 

To sum up, this use case has been among the earliest attempts to understand how the joint ISCC design affects edge learning performance. It provides joint SC${}^2$ resource allocation to maximize the convergence rate for a typical wireless sensing-based human motion recognition task and serves as a novel online design framework for task-oriented wireless FL.

\section{Task-Oriented ISCC for Wireless Edge Inference}\label{sec: Task-Oriented ISCC for Wireless EI}

In this section, we discuss the commonly concerned performance metrics in ISCC-assisted edge inference, followed by a concrete use case of human motion recognition, illustrating how ISCC can be designed to account for all the mentioned metrics and the trade-off among them in the context of edge inference.

\subsection{Performance Metrics}
The commonly used performance metrics for ISCC design in edge inference include inference accuracy, inference latency as well as energy consumption as elaborated below.

\begin{itemize}
    \item \textbf{Inference accuracy}: Inference accuracy characterizes how well the trained AI model generalizes on the unseen data, which can be measured by e.g., classification error or mean square error for classification or regression tasks respectively. In general, it is difficult to derive the exact expression of inference accuracy for an arbitrary AI model. Therefore, approximate surrogates, e.g., discriminant gain as elaborated in the sequel, are needed to develop for facilitating the tractable analysis or optimization on the metric.

    \item \textbf{Inference latency} 
    Inference latency measures how fast the AI model can make predictions on unseen data, which is defined as the wall clock time consumed in the processes of sensing, computation, and communication during the inference pipeline. In contrast to that in the FL counterpart, the concerned walk-clock time consists of only one round of forward sensing-computation-communication delay without any iteration rounds. For those mission-critical or delay-sensitive applications, it usually has a stringent latency requirement, e.g., in the millisecond level for auto-driving applications.

    \item \textbf{Energy consumption}: In the context of edge inference, energy consumption in the SC${}^2$ pipeline shares a similar definition as its FL counterpart mentioned earlier. Energy-efficient inference techniques are highly desired to allow the widespread hardware-primitive and energy-limited IoT devices to be substantially empowered by AI inference capability, so as to achieve the grand vision of pervasive intelligence in B5G IoT.

\end{itemize}


\subsection{\emph{Use Case: Multi-view Sensing}}

\begin{figure}[t]
\centering
 \setlength{\abovecaptionskip}{0in}
\setlength{\belowcaptionskip}{0in}
    \includegraphics[width=4.8in]{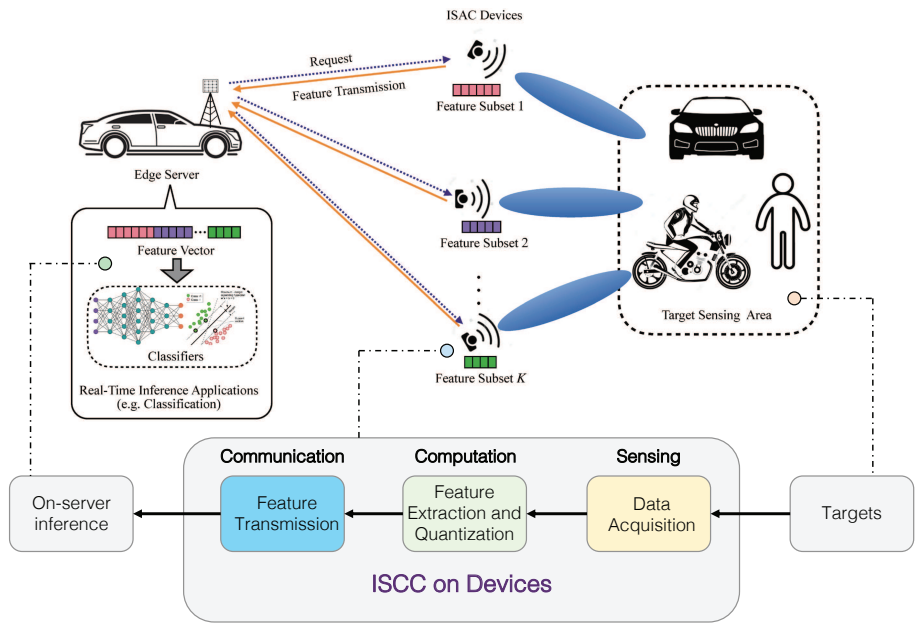}
  \caption{The diagram of ISCC for edge inference.}
  \label{fig:ISCC for inference}
\vspace{-0.1in}
\end{figure}

 The emerging mission-critical and latency-sensitive intelligent applications, e.g., auto-driving and metaverse, call for tasked-oriented technologies that concern the efficient and effective execution of edge inference over wireless networks, in which sensing, communication, and computation are highly coupled ingredients, corresponding to data generalization, exchanging and processing respectively, and thus need to be jointly designed. Specifically, as shown in Fig. \ref{fig:ISCC for inference}, all the three processes mentioned may introduce a certain amount of distortion to the features uploaded from the edge devices to the edge server for inference due to e.g., sensing resolution, feature compression, and wireless channel fading and noise in the pipeline. In particular, sensing, computation, and communication compete for the limited on-device resources (e.g., energy and radio resources), and thus the resource allocation among the three processes has a complex impact on the ultimate inference performance, which need to be judiciously optimized.

To better illustrate the task-oriented ISCC design principle, we consider a multi-view ISAC-based edge split inference system for human motion recognition \cite{Wen-TWC2022}. \textcolor{black}{In this system, multiple edge ISAC devices perform wireless sensing to obtain multi-view sensing data for a common target area (for, e.g., environment surveillance purpose), and then offload the quantized version of extracted features to a centralized edge server, which then conducts the model inference based on the cascaded feature vectors as depicted in Fig. \ref{fig:ISCC for inference}. The objective of this system is to attain high-accuracy inference in real-time so as to meet the stringent delay and reliability requirement of emerging mission-critical applications e.g., auto-driving or metaverse.
The interplay among the SCC processes in this concrete use case lies in that the sensing and communication therein compete for the radio resources (e.g., time and energy) due to the adoption of the ISAC technique, and the allowed communication resource further determines the required quantization level such that the quantized features can be transmitted reliably to the edge server under the given delay and energy constraints. That said, higher sensing quality comes at a cost of severer quantization distortion for reducing communication bits to meet the stringent resource constraints and vice verse.} Such a tradeoff in ISCC has been mathematically cast as an inference accuracy maximization problem with sensing and communication power, time, and quantization bits to be optimized subject to on-device energy and latency constraints.
Nevertheless, the first challenge encountered in formulating the ISCC problem for edge inference is the lack of tractable measures for inference accuracy. To overcome this, the authors in \cite{Wen-TWC2022} proposed and derived an approximate surrogate metric called discriminant gain that is defined as the centroid distance between two classes in the Euclidean feature space under normalized covariance. Most importantly, the surrogate metric allows us to quantify and have a geometrical understanding of how the feature distortion incurred in the sensing, computation, and communication processes affect the inference accuracy, thus shedding light on the subsequent resource allocation. Remarkably, despite the non-convexity of the problem even with a tractable inference accuracy measure, the problem was optimally solved based on the advanced \emph{sum of ratios} method \cite{Wen-TWC2022}. The significant performance gain of the proposed ISCC design over the classic separate SCC design can be observed as evident by the extensive experiments provided in \cite{Wen-TWC2022} using a high-fidelity wireless sensing simulator for human motion recognition.  

In a nutshell, this use case represents the very first attempts to characterize the interplay among SCC processes and their combined effect on the ultimate system performance in the context of edge inference. At a higher level, the task-oriented ISCC resource allocation frameworks established in this use case and those in the previous section serve as two key examples to inspire more follow-up research on developing task-oriented technology for edge AI.

\section{Advanced Task-Oriented ISCC Techniques for Wireless Edge AI}\label{sec: Advanced Task-Oriented ISCC Techniques}

\subsection{Scalable Task-Oriented ISCC}\label{subsec: ICC for FL over Wireless D2D Networks}
\textcolor{black}{With the proliferation of geo-distributed IoT devices with siloed data, distributed learning based on PS-client architecture alone cannot satisfy growing demand for massive ISCC tasks with stringent requirement on end-to-end latency, due to coverage or feasibility of an access point (AP), prolonged transmission delay and limited tolerance to failure of such centralized orchestrators. To address this issue, decentralized architecture for wireless FL relying on peer-to-peer communication topology without a PS has emerged, where a group of edge devices collaborates to train a shared model over wireless device-to-device (D2D) networks \cite{HXing2021Ar}. As an example, in smart transport settings, intelligent vehicles (IVs) equipped with various sorts of sensing units (camera, radar and LIDAR, etc.) collect their sensing information about environment as input data to collaboratively train a target detection model or reinforcement learning (RL)-assisted autonomous decision making, while periodically agreeing on their local model by vehicle-to-vehicle (V2V) communications over a fully self-organizing network.}


\begin{figure}[t]
	\centering
	\setlength{\abovecaptionskip}{0in}
	\setlength{\belowcaptionskip}{0in}
	\includegraphics[width=5.0in]{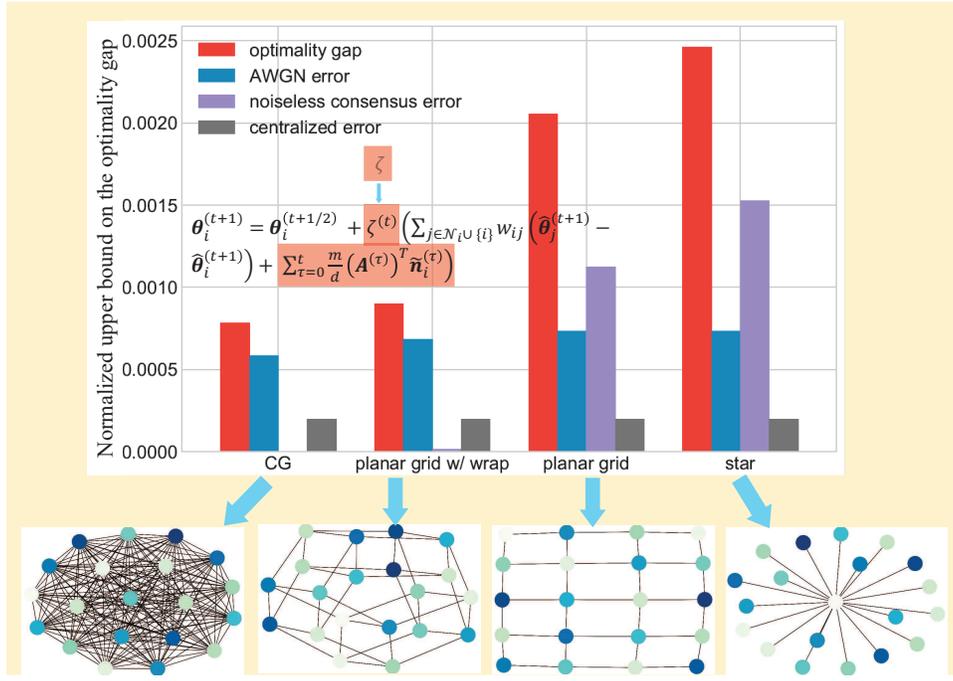}
	\caption{Decomposition of upper bounds on the optimality gap for different D2D topology: 1) CG, 2) planar grid with wrapping, 3) planar grid (without wrapping) and 4) star of the underlying connectivity \cite{HXing2021Ar}.}
	\label{fig:impact of connectivity on AirComp over D2D}
	\vspace{-0.1in}
\end{figure}
One classical type of training algorithm adopted for reducing communication overhead in these decentralized setups is \emph{CHOCO-SGD} \cite{koloskova19decentralized}. The work in \cite{HXing2021Ar} exemplified how computation and communication can be integrated into analog wireless implementation of CHOCO-SGD to enable communication-efficient exploit of channel uses, while providing theoretical insights into the impact of wireless hostilities, e.g., link blockages, added white Gaussian noise (AWGN) and fading etc., and the underlying network connectivity on convergence performance. In each communication round of CHOCO-SGD, communication blocks are divided into equal-length slots, and these slots are scheduled in pairs to enable AirComp. Specifically, in the first slot of the pair, multiple center devices receive in a parallel superposition of signals transmitted by their respective neighbors; and in the second slot of the pair, these center devices communicate back to the associated neighbors by broadcasting. This AirComp-based decentralized FL has therefore two major differences from ideal communication-based FL as follows. First, for each device, the estimate of the combined model parameters from all its neighbors is obtained in multiple slots depending on the device-scheduling scheme. Second, the actual consensus update serves as a noisy approximation to that in ideal communication due to accumulated channel noise added to it as shown in Fig. \ref{fig:impact of connectivity on AirComp over D2D}, thus negatively affecting the upper bound on the optimality gap of the wireless FL. Fig. \ref{fig:impact of connectivity on AirComp over D2D} also shows that the approximation error caused by the AWGN remains severe even over well-connected graphs such as complete graphs (CGs) and planar grids.

\textcolor{black}{In spite of benefiting from scalability and communication efficiency enabled by D2D-based AirComp-FL, it remains challenging to mitigate the effect of accumulated AWGN due to analog transmissions. Possible solutions include joint algorithmic design of adaptive consensus rate (c.f.~$\zeta^{(t)}$ in Fig. \ref{fig:impact of connectivity on AirComp over D2D}) and power control policies to minimize such deterioration to the learning performance, and device sampling to identify only necessary local model aggregation. However asynchronous consensus updates incurred by device sampling pose new challenges to convergence of decentralized training algorithms. Also, the accumulated AWGN gauges additional layer of data privacy on top of FL as long as its level is under control, which will be further discussed in the next subsection.}

\subsection{Privacy-Aware Task-Oriented ISCC}
The success of ISCC relies on massive data acquisition, however, incurs the risk of privacy leakage. The privacy issue is more critical to edge inference where the raw data is processed by ISCC. The sensed raw data may reveal personal information during communication or computation out of the devices. As an example in health care, the sensed data includes location and proximity for tracking the spread of diseases or physiological data such as blood pressure and glucose as indicators of a person's health status. 
In the context of edge learning, the FL paradigm is in favor of privacy preservation by sharing model updates rather than raw data. Nevertheless, the malicious third party can still infer the existence of personal data from the shared information by membership inference attack  or model inversion attack.

\begin{figure}[t]
\centering
 \setlength{\abovecaptionskip}{0in}
\setlength{\belowcaptionskip}{0in}
    \includegraphics[width=0.8\linewidth]{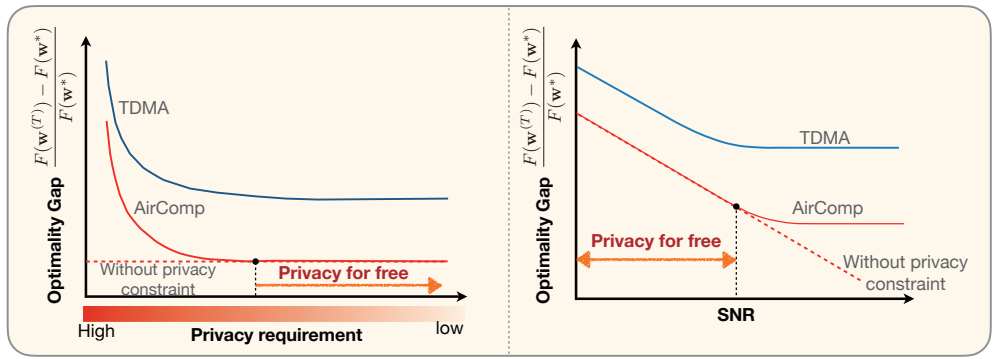}
  \caption{Illustration of the convergence behavior by taking into account the privacy constraint in AirComp-enabled FL systems \cite{liu2020privacy}.}
  \label{fig: sim_privacy}
\vspace{-0.1in}
\end{figure}
Differential privacy (DP) is a formal measure for quantifying information leaked about individual data points. More specifically, it measures the sensitivity of the disclosed statistics to changes in the input data set at a single data point. The classic method to guarantee DP is to introduce a level of uncertainty, e.g., random noise, into the shared information that is sufficient to mask the changes of any individual data point. As a tentative example of privacy-aware ISCC, we consider the AirComp-enabled FL where privacy is attained by channel noise that is added to the uncoded transmitted signals \cite{liu2020privacy}.  Adapting the transmitted power can attain different privacy levels while affecting the convergence speed. As shown in Fig. \ref{fig: sim_privacy},  under a less stringent privacy requirement or lower SNR regime, this design can guarantee privacy without compromising the learning performance, that is, the privacy is obtained for free. AirComp can further enhance privacy as a benefit of multi-user diversity gain. On the one hand, the change from an individual data point can be masked by the superposition of the signals over the air, which enhances the privacy of each individual user. On the other hand, multiple users reuse the channel noise from the non-orthogonal MAC, thus achieving multiple DP mechanisms simultaneously.

In a more general sense, we should exploit the randomness that is inherent in ISCC due to the electronic components or communication environments. Mitigating the randomness for a more reliable ISCC is unnecessary in edge AI, especially under the privacy requirement.  As per DP, a more random system can better preserve privacy. However, increasing the randomness is not monotonically enhancing the edge AI performance. An adequate amount of randomness can be beneficial, e.g., escaping the saddle point for improving the loss function or avoiding overfitting for a better generalization capability, while the overuse can be harmful. This motivates rethinking the theoretical tradeoff between privacy and edge AI performance and designing privacy-aware ISCC to be in the regime that achieves the alignment of both. Another approach to alleviating the privacy risk is to reduce the amount of data to be exposed. The data source is generated by sensing and thus motivates the idea of sensing when necessary. The key reason behind this is that not all data samples are important for improving task performance.  A similar principle has been applied to reduce the communication cost where only a few data samples are selected for transmission according to the uncertainty --- a measure to quantify the improvement for the learning accuracy \cite{liu2020wireless}. Accordingly, a tractable approach for privacy-aware ISCC is to use the current task performance and privacy requirement as feedback to control the sensing frequency, and thus, adapt the amount of data to be used in edge AI. Reducing the data usage synergistically alleviates the communication and computation costs toward the efficient ISCC.

\subsection{Energy-Efficient Task-Oriented ISCC}
Edge AI applications can be a serious energy drain on smart edge devices. For instance, an edge processor equipped with a $2.1$\,Wh battery can only run a \textcolor{black}{VGG-E network} for about $25$ minutes \cite{roy2019towards}. In this subsection, we introduce the state-of-the-art event-driven SNN-based neuromorphic communications (NeuroComm) proposed in  \cite{chen2023neuromorphic} in order to facilitate the development of energy-efficient task-oriented ISCC. 

NeuroComm can be regarded as an SNN-based end-to-end ISCC system, in which the coded sensed signal is a timing of spikes and thus can be directly fed to SNNs to perform inference. Moreover, in NeuroComm, the energy consumption of communication and computation is a function of the activity of the sensed signal. If the sensor produces no spikes, the SNN is idle and there is no signal to be transmitted, i.e., the communication energy consumption is zero. This enables NeuroComm to be highly energy efficient and event-driven. Moreover, supported by the large-scale neuromorphic chips, each spike in on-chip SNNs only costs a few picojoules.

\begin{figure}[t]
  \centering
   \setlength{\abovecaptionskip}{0in}
  \setlength{\belowcaptionskip}{0in}
      \includegraphics[width=0.8\linewidth]{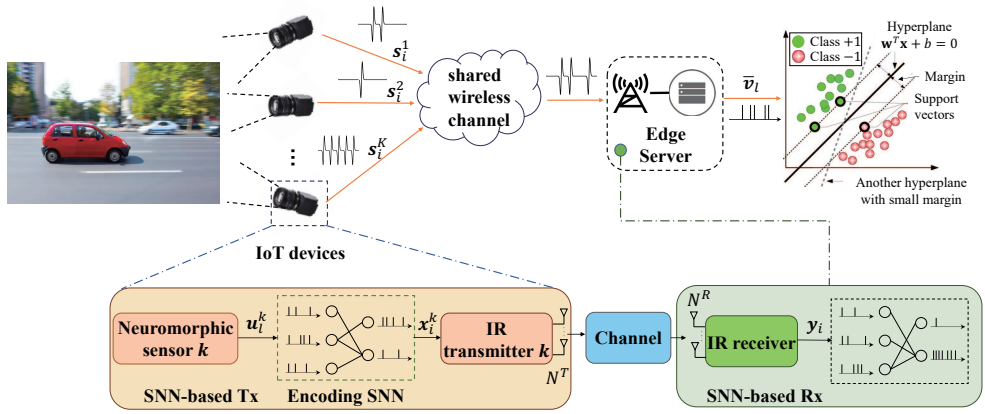}
    \caption{The diagram of NeuroComm systems \cite{chen2023neuromorphic}.}
    \label{fig:NeuroComm_system}
  \vspace{-0.1in}
  \end{figure}
In the NeuroComm system as shown in Fig. \ref{fig:NeuroComm_system}, neuromorphic sensors, SNNs, and impulse radio (IR) transceivers are integrated. The neuromorphic sensor $k$ captures environmental information and encodes it as a spike timing sequence $\mathbf{u}_{l}^k$. For instance, dynamic vision sensors (DVS) produce spikes when pixel values change significantly. The SNN-based encoder processes $\mathbf{u}_{l}^k$ and generates the encoded sequence $\mathbf{x}_{i}^k$, which is transmitted using IR transceivers employing time-hopping modulation. In this way, the communication energy consumption depends on the activity of the sensed signal and the encoding SNN output. Then, the receiver decodes the received timing of spikes $\mathbf{y}_i$ using the SNN decoder and performs inference. However, direct use of the received signal $\mathbf{y}_i$ for inference is challenging due to channel corruption. Previous methods use pilots for channel equalization but lack task-oriented design. The work in \cite{chen2023neuromorphic} utilizes a hyper-SNN, where channel coefficients are incorporated into training and inference. The hyper-SNN maps the received pilots to the weights of the decoding SNN, while the weights of the encoding SNN remain fixed. Finally, the entire system is trained in an end-to-end fashion, focusing on task-oriented loss, e.g., e.g., cross-entropy for classification, instead of reconstruction error.

Moving forward, there are several open issues in task-oriented energy-efficient ISCC. One area of development is the creation of transfer learning algorithms to adapt pre-trained SNNs to new ISCC tasks. Additionally, designing high-efficient input/output interfaces between SNNs and other modules (e.g., neuromorphic sensors and IR transceivers) holds importance in further enhancing energy efficiency.

\section{Conclusions and Discussions}\label{sec: Conclusions}
\textcolor{black}{This article provided an overview of principles of enabling ISCC technologies, including ISAC, ICAC, and ISA``C'', followed by two concrete use cases of edge AI tasks to demonstrate the advantages of task-oriented ISCC. Furthermore, in scenarios where PS-client architecture is not available due to coverage, feasibility or one-node failure, where private data is exposed to malicious inference in spite of FL settings, or where low-energy IoT devices struggle to afford DNN inference, we have identified some advanced techniques including decentralized FL over wireless D2D networks, adaptive power control for DP in AirComp-based FL, and energy-efficient NeuroComm for event-driven edge inference to address them respectively.}

\textcolor{black}{Finally, there still remain many open challenges in the implementation of these advanced techniques worth of further investigation. 
For instance, IoT devices in decentralized wireless sensor networks can pose drastic heterogeneity in terms of data-sample quality, local computation and communication capacity in one round of training, which causes unbearable delay if one device awaits model aggregation from a straggling neighbor, or very low learning efficiency if the sensed data is too poor to extract information. One promising solution is to adopt device sampling that accounts for both statistical and system heterogeneity, along with asynchronous decentralized training algorithms that ensure convergence under dynamic connectivity. In addition, when deploying AirComp-based ISCC solutions, synchronization of multiple access to the PS for waveform superposition and accurate channel estimation for signal alignment are two major challenges. Possible solutions encompass estimation of timing offset (TO) and carrier frequency offset (CFO), followed by pre-equalization prior to analog transmission, and power control policies aimed at minimizing the overall transmission deterioration to learning performance with robustness against channel estimation error.
}

\bibliography{AirComp4FL}
\bibliographystyle{IEEEtran}

\end{document}